\documentclass[final,3p,times,12pt,]{elsarticle}
\usepackage{amssymb}

\usepackage{graphicx}
\journal{--arxiv-----}

\begin{document}
\begin{frontmatter}
\title{ Lunar eclipse induces disturbance in the lunar exosphere }

\author[1]{Anil Raghav\corref{cor2}\fnref{fn1}}
\ead{raghavanil1984@gmail.com /anil.raghav@physics.mu.ac.in}
\cortext[cor2]{Principal corresponding author}
\fntext[fn1]{ Present address: Department of Physics, University of Mumbai, Vidyanagari, Santacruz (E), Mumbai-400098, }
\author[2]{Ankush Bhaskar}
\author[2] {Virendra Yadav}
\author[1]{Nitinkumar Bijewar}
\author[1]{Chintamani Pai}
\author[1]{Vaibhav Rawoot}
\address[1]{Department of Physics, University of Mumbai,Vidyanagari, Santacruz (E), Mumbai-400098, India}
\address[2]{Indian Institute of Geomagnetism, Kalamboli Highway, New Panvel, Navi Mumbai- 410218, India.}

\begin{abstract}

{Given the renewed scientific interest in lunar exploration missions, complete understanding of lunar near surface environment and its exosphere under different conditions is of paramount importance. Lunar exosphere has been extensively studied by ground based observations\cite{18,19,20,21,22,23} and hypothesized by different models\cite{1,2,3,4,5,6,7,8,9,10,11,12,13}. In present work, we have discussed overlooked possible sources behind changes in the lunar exosphere when the Moon passes through the penumbra and umbra of the Earth during a lunar eclipse. The dusty turbulent environment due to planetary shadow is not only confined to lunar studies and exploration, but it can also be extended to all terrestrial airless bodies in the universe with a dusty surface e.g. some planets, planetary satellites, asteroids etc.
}

\end{abstract}


\end{frontmatter}

\section{main}
\label{1}
{

The lunar surface is rich in topographic features and covered with very active regolith, 
creating a dynamic dusty near-surface environment and extended tenuous exosphere. 
This consists of a broad distribution of particle size and has been hypothesized on
the basis of ‘‘horizon glow’’ observations during the Apollo era \cite{1,2,3,4,5}. 
The extent and dynamics of the lunar exosphere has been extensively studied and probed
by satellites and ground based observations since 1970s. Attempts have been made to explain
this dusty near-surface environment and horizon glow of the Moon by various electrostatic 
lofting models, also applied to airless celestial bodies \cite{6,7,8,9,10,11,12}. Models 
suggest that the key factor involved in lunar dust dynamics is the complex electric field
near the surface. The lunar surface and the dust particles are charged by their electrostatic
interaction with the local plasma environment. Consequently, negative potential develops on 
the night side of the Moon.  On the day side positive potential is developed due to photo-emission
of electrons by interaction of solar UV and X-rays. This electrostatic interaction creates like 
charges on the surface and on the dust particles due to which they repel each other. The most 
complicated electric field region on the Moon is near day-night terminator, where there is a 
distinct transition from the sunlight-driven positive to the plasma-driven negative surface 
potential \cite{8,10,11,12}. Similar intense horizontal and vertical electric fields have 
been observed near surface in the laboratory model of differential UV charging of the lunar surface\cite{13}.

This complex vertical and horizontal electric fields have been incorporated in the ‘fountain model’ of dust 
dynamics which explain ejection mechanism of dust at the day night terminator \cite{6,7}.  The transition 
from the maximum positive surface potential to a negative one implies the existence of ‘dead zone’, a 
location where the surface potential is zero.  The dead zone has low dust activity whereas the terminator
is a high dust activity region. The ``dynamic dust grain fountain model'', which assumes one order larger 
potential on the lunar dark side as compared to the lit side, accounts for lofting of $~$ 0.01 $–$ 0.1µm 
dust grains to altitudes $~$ 0.1 – 100 km near the terminator. It predicts that smaller size dust grains
can be lofted to higher altitudes and can move with velocities greater than the lunar escape velocity. 
The detailed physical mechanism behind this dust ejection has been discussed in past studies \cite{6,7}.

In this paper we are proposing that a lunar eclipse induces an effect similar to the day-night terminator
effect on the lunar day side, which we have termed as Eclipse Induced Terminator (EIT). It is a boundary
which separates the illuminated and eclipsed regions of the lunar surface. The rapid EIT progression during
a lunar eclipse across the day side surface creates regions of differential charging. This leads to vertical
uplift and horizontal transport of the charged lunar dust creating dust fountain along the EIT. The assumption
that EIT is similar to the day-night terminator can be validated on the basis of the particle experiments done
by the University of California, San Diego (UCSD) on the geosynchronous satellites ATS 5 and ATS 6. These 
experiments observed large potentials during the passage of the satellites through the Earth's shadow \cite{14}.
DeForest and Mcllwain have interpreted this as the photoelectron flux going to zero shifting the already negative
potential even more negative when the satellites entered the shadow \cite{15}. Potential drops by one order were 
observed when the satellites entered the shadow from sunlight.  This potential drop is similar to the potential
difference between the lunar day and night regions discussed in the fountain model. However, any detailed treatment
of EIT phenomenon will also have to include the surface temperature gradient during a lunar eclipse which has been
observed in lunar surface radiation measurements. Temperature drops from 371K to 200K in partial phase and to 175K
during totality have been observed \cite{17}.  This rapid temperature variation can change the material surface 
properties which can modulate the dust dynamics in the lunar environment.  The dust particles which are away from
the EIT, experience weaker electric field and try to settle down under gravity. The continuous horizontal and 
vertical transport and the tendency of the dust to settle down under gravity result in a dynamic dusty 
environment on the lunar day-side during the eclipse. The scenario is even more complex given the fact that
there exist different plasma properties in the Earth's umbra and penumbra and their boundaries\cite{14,15,16}.
All this will affect the lunar exosphere parameters.

\begin{figure}
\centering
\includegraphics[angle=0,width=80mm]{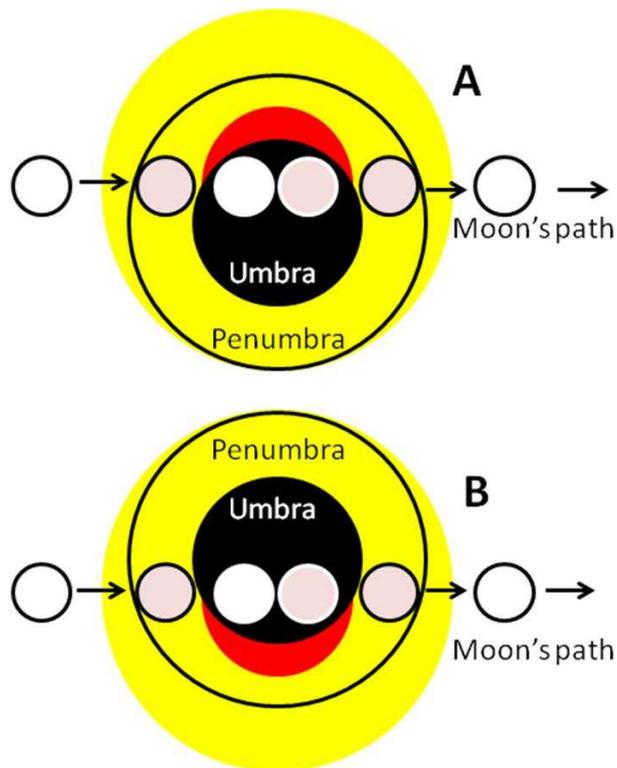}
\caption{Schematic diagram of the asymmetry in sodium brightness distribution. Red and yellow represent higher and lower concentrations respectively. Higher concentration is observed in the direction of the nearest penumbral region.}
\end{figure}

Many spectroscopic studies of the lunar exosphere have been carried out \cite{18,19,20,21,22,23,24}. 
Spectroscopic studies during lunar eclipses provides confirmation of lunar Na exosphere \cite{20} and its full extent to distances of $~$20 RM \cite{23}.
These studies concluded that the thermal desorption, 
in which heating of the surface evaporates atoms and molecules off of the surface; photon-stimulated desorption, 
which involves an electronic excitation of a target atom or molecule; sputtering by solar wind ions, which ejects
atoms or molecules by either a cascade of momentum transfer collisions or chemical reactions which reduce the 
binding energies within the surface material; and meteorite impact vaporization might involve in observed lunar exosphere. 
Spectroscopic studies suggest most of the Na atoms captured in  field of view are escaping from the Moon, and thus lunar eclipse
observations offer unambiguous evidence of mechanisms responsible for the high-speed component of the Moon’s
surface-bounded exosphere \cite{23}. Mendillo and Baumgardner(1995), eclipse study indicate some process 
other than solar wind sputtering or solar wind ESD was contributing to the extend of exosphere \cite{21}. Even-though 
obvious geometrical and temporal parameters test such as Solar zenith angle tests, Meteor shower tests, 
 and  Solar flare tests considered to understand the extension of lunar exosphere.
Wilson (2006) et.al conclude that the higher UV fluxes in November 1993 and July 2000 did not
significantly affect the exospheres observed during eclipse. Since, the Moon was shielded from the solar wind and 
there were no major meteor showers during any of the observations that eliminates the cause of solar wind and meteor impact as well.
they also look for the changes in the magneto-tail might somehow account for the difference between
eclipse exospheres \cite{22}. But, no conclusive cause obtain regarding the extension of lunar exosphere.


 However, this extended lunar exosphere is of direct consequence of the hypothesis given above. It can be seen that the traversal path of the Moon in the umbra and penumbra region is different in different eclipses observed. In each of the eclipses, the observed distribution of sodium brightness is different and always follows the direction of the nearest penumbral region. The schematic diagram of the sodium brightness distribution with respect to the Moon’s passage through umbra and penumbra region is shown in fig.1. That means the different plasma conditions in the umbra and penumbra along with EIT, which provides the ejection mechanism for the lunar dust, is responsible for the changes observed in the lunar exosphere during a lunar eclipse. On the similar lines, we can also expect changes in the lunar exosphere when the Moon passes through the Earth’s magnetotail, where it encounters plasma sheet \cite{22}.
 
The predictive validity of hypothesis can be established using lunar gamma ray albedo measurements or satellite measurements. Till now, the models and the experimental findings of shadow effect and gamma-ray albedo have considered the Moon as a solid body, devoid of any interacting atmosphere. The GCR flux incident on the Moon at small zenith angles goes deep into the surface and produces low energy secondary cosmic ray flux. Most of the high energy secondary cosmic rays are produced by the GCR hitting the lunar surface tangentially\cite{25,26,27}. This indicates dusty lunar environment might significantly contribute in the high energy secondary cosmic ray production. During a lunar eclipse, dynamic dusty conditions increase the dust density of the exosphere as compared to that during an un-eclipsed full moon, thereby creating additional interaction centers for GCR. This should lead to increase in secondary cosmic ray flux in the lunar environment. This secondary cosmic ray flux can reach the surface of the Earth by going through multiple interactions with the atmosphere. In a steady atmosphere, part of this should be reflected in the ground based observations of secondary cosmic ray flux, provided solar and geomagnetic conditions are quite. The measurements of three lunar eclipses presented by Anand Rao (1967) and lunar eclipse measurement done by A. Raghav et. al. shows enhancement in secondary cosmic ray flux\cite{28,29}. This might be a signature of the above discussed predictions.  A detailed discussion of the results presented by A. Raghav et al. with respect to the above proposed hypothesis is incorporated in appendix.

This hypothesis proposes new possible sources of change in the lunar exosphere during a lunar eclipse. Also, the generalized assumption made for spectroscopic observations of sodium brightness distribution, that the conditions during a full moon and a lunar eclipse are same is not completely valid. This has immense implications for lunar missions due to the known hazards arising from the complex electric field and dust\cite{30}. The dusty turbulent environment due to planetary shadow is not only confined to lunar studies and exploration, but it can also be extended to all terrestrial airless bodies in the universe with a dusty surface e.g. some planets, planetary satellites, asteroids etc. Further studies are required for the comprehensive understanding of this phenomenon.

\section{Appendix}

\title{Enhancement observed in ground base observation of secondary cosmic ray flux measurements during four different lunar eclipses}

The presented figure is adopted from A. Raghav et al., 2012 who reported observations of the  secondary cosmic rays during the lunar eclipse.  Here we discuss their reported observations with respect to the proposed physical scenario.

As seen in the above figure approximately 8.3$\%$ secondary cosmic ray flux enhancement during the lunar eclipse with respect to the average pre- and post-eclipse counts. To explain this enhancement they studied various factors affecting the secondary cosmic ray flux like atmospheric variations, geomagnetic variations and lunar tidal effects. The local atmospheric parameters like temperature, humidity, pressure etc. are expected to be independent of lunar eclipse. The recorded temperature and relative humidity data do not show any correlation with the observed enhancement in the secondary cosmic ray flux. The Kp index remained between 1-2 during the eclipse which shows quiet geomagnetic condition. Lunar tide is maximum when the Moon is near zenith or nadir, whereas the maximum secondary cosmic ray flux was observed when the Moon was at low elevation. Based on the above arguments they emphasized that the observed enhancement in the profile cannot be explained on the basis of the tidal effect. This led us to consider local electromagnetic field disturbances near the lunar surface as a probable cause of the secondary cosmic ray flux enhancement observed during the eclipse.

\begin{figure}
\centering
\includegraphics[angle=0,width=80mm]{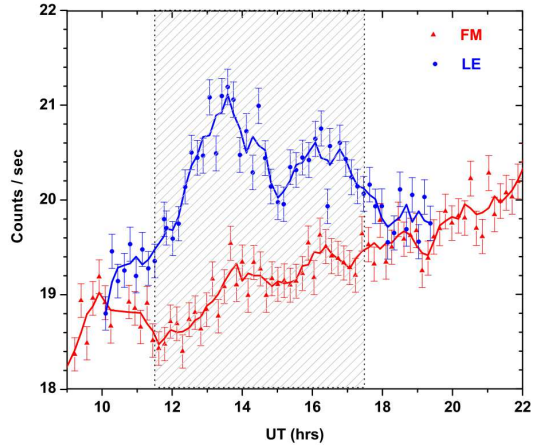}
\caption{SCR variation during the lunar eclipse (LE) and full moon (FM) recorded by scintillation detector. The dotted lines indicate respective contact timings of the Moon with the Earth's penumbra (P) and umbra (U). Central black dotted line indicates the greatest eclipse time.}
\end{figure}

The observed enhancement in the secondary cosmic ray flux coincides well with the lunar entry in and exit from the Earth's penumbra at P1 and P4 respectively, which indicates that this enhancement is related to the Earth’s shadow. We suggest that this may be a signature of the complex electric field generated dust storm on the Moon as proposed in paper. As the Moon starts entering the Earth's penumbra, a condition of differential photoelectron charging is established due to the uneven exposure of the lunar surface to sunlight. This creates complex electric field disturbances near the lunar surface along EIT which causes a dust storm and leads to a gradual increase in the secondary cosmic ray flux from P1 to U1. The maximum flux is observed when the Moon is in between U1 and U2 that means largest EIT length which we propose to be the condition of maximum dust availability for interaction with the GCR flux. The gradual decrease in the GCR flux thereafter till U3 indicates a decrease in the EIT induced complex electric field region and settling of the lifted dust. One expects the reverse effect to be observed during the lunar exit from U3 to P4. However, the conditions are not exactly the same as before since the direction of the horizontal component of the complex electric field reverses and the availability of dust near the surface is less due to the previous upliftment of dust. Therefore the enhancement in secondary cosmic ray flux during the lunar exit from umbra (U3) is not as pronounced as before. 
Similar enhancement observed during three lunar eclipses by Anand Rao which he presented in Physics Letters A 25, 74 (1967).

\section*{Acknowledgments}

{
We are thankful to the Department of Physics, University of Mumbai, Mumbai, India for providing the experimental resources and facilities. We are thankful to the Department of Chemistry, University of Mumbai, Mumbai, India for providing lead shield during experiment. We take a great pleasure in expressing our gratitude to S. M. Chitre, S. K. Tandel and S. B. Patel of UM-DAE Center for Excellence in Basic Sciences for their constant encouragement and support. We are extremely grateful to D.C. Kothari, M. R. Press, A. Misra, A. Patwardhan, C. V. Gurada and R. Srinivasan of the Department of Physics. We thank G. Vichare and D. Tiwari of Indian Institute of Geomagnetism, Navi Mumbai. We also thank  N. Dubey, A. Koli, N. Navale, G. Palsingh, P. Parab, G. Narvankar, S. Chalke, N. Malandkar, S. Pawar, S. Borse, S. Tiwari, P. Yadav. We extend our gratitude to S. Kasturirangan, D. Misra, J. More, R. Kadrekar, S. Kadam, S. Sathian and A. Johri.
}
\end{document}